\documentclass[aps,prb,twocolumn,superscriptaddress,showpacs,10pt]{revtex4-1}
\usepackage{natbib}
\usepackage{mathbbol,amsmath,amsfonts,amssymb,bm,color,dsfont,bbold}
\usepackage{graphicx,psfrag,array,braket}

\providecommand{\ignore}[1]{}
\providecommand{\aucmnt}[1]{#1}
\def\be{\begin{equation}}
\def\ee{\end{equation}}
\renewcommand{\aucmnt}[1]{}
\newcommand{\Comment}[1]{}

\begin{document}
\title{Construction of a series of new $\nu=2/5$ fractional quantum Hall  wave functions by conformal field theory}

\author{Li Chen}
\affiliation{College of Physics and Electronic Science, Hubei Normal University, Huangshi 435002, China}
\author{Kun Yang}
\affiliation{National High Magnetic Field Laboratory and Department of Physics, Florida State University, Tallahassee, FL 32306, USA}

\date{\today}
\begin{abstract}
In this paper, a series of $\nu=2/5$ fractional quantum Hall  wave functions are constructed from conformal field theory(CFT). They share the same topological properties with  states constructed by Jain's composite fermion approach. Upon exact lowest Landau level(LLL) projection,  some of  Jain composite fermion states would not survive if constraints on Landau level indices given in the  appendices of this paper were not satisfied. By contrast, states constructed from CFT always stay in  LLL. These states are characterized by different  topological shifts and multibody relative angular momenta. As a by-product, in the  appendices we prove the necessary conditions for general $ \nu=p/(2p+1) $ composite fermion states to have nonvanishing  LLL projection.
\end{abstract}
\maketitle

\section{Introduction}
Strongly correlated electron system has long been a topic of focus in condensed matter physics, especially the fractional quantum Hall system of electrons on a 2D surface, at temperature as low as several kelvins and in a high magnetic field of several teslas. This system exhibits fractional quantum Hall effect, in which Hall resistance $R_H$ is quantized as ${\frac{h}{{\nu {e^2}}}}$ in a certain range of magnetic field, with Planck constant $h$, electron charge $e$ and fractional filling factor $\nu$. Fractional quantum Hall effect, as well as integer quantum Hall effect in which filling factor $\nu$ is an integer, can be explained by the notion of Landau level(LL).
Under the above experimental  circumstances, energy spectra for a single electron  will form discrete energy levels known as LLs. Completely filled LLs account for the integer quantum Hall effect while partially  filled LLs with a certain  fractional  filling factor $\nu$ lead to the fractional quantum Hall effect, where $\nu=N_e/N_{\Phi}$. Here $N_e$ is electron number and $N_{\Phi}$  is LL degeneracy, which is equal to the number of flux quanta piercing the system.  The most well known fractional quantum Hall effect is the one with $\nu=1/3$, where onethird of the lowest Landau level(LLL) is filled. In the regime of fractional quantum Hall effect, if we project the full Hamiltonian of 2D electron system in a high magnetic field onto LLL, the single-particle kinetic energy term is quenched while interaction  terms remain. Still this Hamiltonian cannot be easily  solved, so physicists resort to trial wave functions, such as the Laughlin wave function(which was proposed by Robert Laughlin \cite{Laughlin}) for the filling factor $ 1/3 $ and Jain's composite fermion wave function for the series with filling factors $p/(2p+1)$.
Model Hamiltonians can be derived from these trial wave functions, which are easier to deal with on analytical grounds than most realistic Hamiltonians. When projected onto specific LLs of interest, these Hamiltonians usually assume forms of 1D frustration-free lattice Hamiltonians which, in their second-quantized forms, have been studied thoroughly, and many properties of their zero modes(zero-energy ground states) have been discovered \cite{ortiz,Chen17}.
On the other hand, trial wave functions for the fractional quantum Hall  effect have been connected to conformal field theory(CFT). The Laughlin wave function, as an example, can be cast as the conformal correlator of massless free boson in CFT \cite{Fubini,LaughlinCFT}. Later on, new wave functions were proposed from conformal correlators, such as the famous Moore-Read Pfaffian and Read-Rezayi wave functions, argued to possess quasiparticle/quasihole excitation obeying non-Abelian anyonic statistics \cite{MR,RR}. The justification for the construction of  trial wave functions from CFT is that the boundary theory of Chern-Simons theory which characterizes the quantum Hall effect is a CFT \cite{RevModPhys.80.1083}, and the ground-state wave function can be viewed as the amplitude of particle configuration in a time slice of such a 2+1 D system.

It took a few more years for people to realize that Jain's composite fermion wave functions, such as that for $\nu=2/5$ , once projected to LLL, is also a conformal correlator \cite{CFTJain1,CFTJain2}. Since Jain's $\nu=2/5$ composite fermion wave function (which is our primary example for Jain states) has two degrees of freedom associated with the two lowest LLs for the composite fermion (more details will be presented in Sec. \ref{jaincf}), its corresponding conformal correlator is constructed from two independent massless free bosons. However unlike the Read-Rezayi sequence which includes the Laughlin and Moore-Read states, constructing the Jain states involves not only the primary, but also descendant fields of the corresponding CFT. Since there is a large degree of freedom in the choice of the latter even when $\nu$ is fixed, there should be a corresponding {\em family} of Jain states.
Motivated by the developments and consideration mentioned above, we have generalized Jain's $\nu=2/5$ composite fermion wave function to cases in which the composite fermions occupy higher LLs, while the lower one(s) may be empty. Since LLL projection is performed, they represent distinct yet legitimate LLL states at the same filling factor. As we are going to show, they correspond to different choices of descendant fields in the CFT construction. Upon LLL projection, some of  composite fermion wave functions will vanish. Nonetheless, the corresponding conformal correlators, which are holomorphic and thus reside in the LLL by construction, are nonvanishing. Most importantly, we find that while these families of Jain states share the same $K$ matrix with the original one constructed by Jain, many of them have different topological shifts and other properties. 

The structure of this paper is as follows: In Sec. \ref{jaincf} we introduce Jain's composite fermion  approach and trial wave functions in this approach, known as composite fermion wave functions. In Sec. \ref{CFTcons}, we introduce the conformal field theoretical construction of composite fermion  wave functions.
In Sec. \ref{gcft},  we   propose  a series of fractional quantum Hall trial wave functions of filling factor $2/5$ constructed from CFT correlators and discuss their connections with general LLL-projected composite fermion wave functions. In Secs. \ref{shf} and \ref{mb}, we characterize them by two criteria: one is topological shift and the other is multibody relative angular momentum. In Sec. \ref{thrsvn}, we discuss two interesting states found in the process of studying parent Hamiltonians in the form of projection operators. One of them has filling factor $\nu$=3/7, the other has $\nu$ very close to 1/2  for  a finite number of particles. 
In the  appendices, we give the necessary conditions for a general composite fermion state to have nonvanishing LLL projection.
\section{Composite fermion approach to fractional quantum Hall effect}\label{jaincf}

The $p/(2p+1)$ series can be explained in a systematic way in the framework of composite fermion (CF) theory \cite{Jain}. In this theory, two flux quanta of  the magnetic field are attached to each electron. The composite of an electron and two flux quanta, named the composite fermion,   experiences an effective magnetic field $B^*=B-2n\Phi_0$ as opposed to the actual magnetic field $B$, where $n$ is the average electron density and $\Phi_0$ is the elementary flux quantum. In this composite fermion approach, the fractional quantum Hall effect of electrons at filling factor $\nu=p/(2p+1)$ with integer $p$ is  mapped to the integer quantum Hall effect of composite fermions at filling factor $\nu=p$. From this approach trial wave functions can be inferred, such as the wave function of the LLL-projected Jain composite fermion state with  filling factor $\nu=2/5$   on a disk \cite{compositeReview}, \be\label{wfcomposite fermion} P_{\text{LLL}}\prod\limits_{i<j}(z_i-z_j)^2 \Psi_2,\ee where $ z=x+i \,y $ is the complex coordinate on the disk, $P_{\text{LLL}} $ projects the wave function onto the lowest Landau level, and $\Psi_2 $ is the $N$-particle wave function  at filling factor $\nu=2$; i.e.,
\textit{two} Landau levels (LLs)  are  filled. The Laughlin-Jastrow factor $\prod_{i<j}(z_i-z_j)^2$ has the effect of attaching two flux quanta to each electron.  The explicit form of $\Psi_2 $ is  \be  \Psi_2=  \left| {\begin{array}{*{20}{c}}
  \eta_{n_1,m_1}(z_1)& \ldots &\eta_{n_1,m_1}(z_N) \\
   \vdots & \ddots & \vdots  \\
 \eta_{n_1,m_{\frac{N}{2}}}(z_1)& \ldots &\eta_{n_1,m_{\frac{N}{2}}}(z_N)\\\eta_{n_2,m_{{\frac{N}{2}}+1}}(z_1)& \ldots &\eta_{n_2,m_{{\frac{N}{2}}+1}}(z_N) \\
    \vdots & \ddots & \vdots  \\
 \eta_{n_2,m_{N}}(z_1)& \ldots &\eta_{n_2,m_{N}}(z_N)
\end{array}} \right|, \ee in which $\eta_{n,m}(z)  $ is the single-particle wave function of   angular momentum $ m\hbar $ in the $ n $th Landau level on the disk by choosing a symmetric gauge for the magnetic field $\mathbf{B}=-B \hat{\mathbf{z}}$. The expression of $\eta_{n,m}(z)  $ is \cite{jainbook}
\be\eta_{n,m}(\mathbf{r})=\frac{( - 1)^n\sqrt {n!} }{{\sqrt {2\pi {2^{m}}(n + m)!} }}z^{m}L_n^m\left(\frac{\bar{z} z}{2}\right)e^{ - \frac{\bar{z} z}{4}},\ee in which the magnetic length $l_B=\sqrt{\hbar/eB} $ has been set to 1 and $ L_n^m (x) $ is the generalized Laguerre polynomial, \be L_n^m (x) =\sum\limits_{i = 0}^{n} {{{(-1)}^i}} {n+m \choose n-i} \frac{{{x^i}}}{{i!}}.  \ee  Note that the maximum power of $\bar{z}$ in $\eta_{n,m}(\mathbf{r})$ is $n$, which will be used in the following sections. For example, single-particle wave functions in 0LL, 1LL(the first excited Landau level) and 2LL(the second excited Landau level) are \be
  \eta_{0,m}(z) = \frac{{z^m}{e^{ - {|z|^2}/4}}}{{\sqrt {2\pi {2^m}m!} }},
\ee
\be
  \eta_{1,m}(z) = \frac{\left( {\bar{z} {z^{m+1}} - 2(m+1)z^m} \right){e^{ - {|z|^2}/4}}}{{\sqrt {2\pi {2^{m+2}}(m+1)!} }}
\ee and \be
\begin{split}
 &\eta_{2,m}(z) = {e^{ - {|z|^2}/4}}\\&\times\frac{({\bar{z}^2} {z^{m+2}}- 4(m+2) \bar{z} {z^{m+1}}+4(m+2)(m+1)z^{m} )}{{\sqrt {2\pi {2^{m+5}}(m+2)!} }}.  \end{split}
\ee Besides the Gaussian factor, the single-particle wave function in 0LL is analytic in $z$,  while that in 1LL and 2LL has $\bar{z}$ and $\bar{z}^2$, respectively.

For $\Psi_2$ in Eq. \ref{wfcomposite fermion}, Jain chose two filled LLs as  LLL and 1LL.
The LLL projection $P_{\text{LLL}} $ is technically accomplished in the following way \cite{jainbook}: we bring all the antiholomorphic coordinates  $ \bar{z}_i $ to the leftmost of the wave function, and then replace them individually with $ 2 \partial_{z_i} $, where the derivative only acts on the polynomial part of the wave function. In Appendix \ref{LLP}, we have also given a closed form for the LLL-projected composite fermion wave function using an alternative approach.

\section{CFT construction of Fractional Quantum Hall WAVE FUNCTION}\label{CFTcons}
The wave function of the LLL-projected Jain composite fermion state of an even number of particles in Eq. \ref{wfcomposite fermion}  can be written as a conformal correlator  in CFT \cite{CFTJain1,CFTJain2,QHCFTRevMod, Kvorning}. In the framework of CFT, we introduce two independent free massless bosonic fields $\phi_1(z)  $ and $\phi_2(z)  $ compactified on two circles of radii $ \sqrt{3} $  and $ \sqrt{15} $, respectively. Their conformal correlator satisfies $ \langle \phi_i(z)\phi_j(w)\rangle=-\delta_{i,j}\ln(z-w) $. Then we introduce two vertex operators, \be V_0(z)=:e^{i\sqrt{3}\phi_1(z)}:\ee and \be\label{des} V_1(z)=:\partial_{z}\left( e^{i\sqrt{\frac{4}{3}}\phi_1(z)}e^{i\sqrt{\frac{5}{3}}\phi_2(z)}\right):.\ee where $ :\quad: $ means normal ordering. Here $V_0(z)$ is a primary field and $V_1(z)$ is a descendant of the primary field $: e^{i\sqrt{\frac{4}{3}}\phi_1(z)}e^{i\sqrt{\frac{5}{3}}\phi_2(z)}:$.
It is easy to see that these two vertex operators represent two species of independent electron operators since $ [V_0(z),V_1(w)]=0 $, $ \{V_0(z),V_0(w) \}=0$ and $ \{V_1(z),V_1(w) \}=0$ hold \cite{CFTJain1,CFTJain2}.
We can define the charge operator as \be Q=\frac{1	}{2\pi i} \oint dz J(z),    \ee
in which the $ U(1) $ current is defined as \be J(z)=\frac{i}{\sqrt{3}}\partial_z \phi_1(z)+\frac{i}{\sqrt{15}}\partial_z \phi_2(z).\ee Then it is trivial to see that \be[Q,V_j(z)]=V_j(z)\ee for $ j=0,1. $ Thus both $ V_0(z) $ and $ V_1(z) $ have the correct electric charge 1 in units of electron charge $e$. The   wave function of the LLL-projected Jain state can be written as a conformal correlator,\begin{widetext}\be \mathcal{A}\{\langle V_1(z_1)V_1(z_2)\cdots V_1(z_{\frac{N}{2}})V_0(z_{\frac{N}{2}+1})V_0(z_{\frac{N}{2}+2})\cdots V_0(z_{N})\rangle \}. \ee  That is, half of the electrons are represented by $ V_0 $ and the remaining half are represented by $ V_1 $. This correlator leads to \be\label{J01} \mathcal{A}\{\partial_{z_1}\partial_{z_2}\cdots \partial_{z_{  \frac{N}{2} }}\prod\limits_{i<j\leq   \frac{N}{2}   }(z_i-z_j)^3 \prod\limits_{ \frac{N}{2} <k<l }(z_k-z_l)^3 \prod\limits_{m\leq  \frac{N}{2}   <n }(z_m-z_n)^2    \} \ee by   using the formula \cite{CFT}\be \begin{split}
\langle :e^{i\alpha_1 \phi(z_1)}::e^{i\alpha_2 \phi(z_2)}:\cdots : e^{i\alpha_N \phi(z_N)}:\rangle =\prod\limits_{i<j} (z_i-z_j)^{\alpha_i\alpha_j}.
\end{split}\ee \end{widetext}For simplicity, in the correlator we have neglected the back ground charge term \cite{MR,CFTJain2}  which accounts for the Gaussian factor.  It has been proved  \cite{CFTJain1,CFTJain2} that the  LLL-projected  wave function  prescribed by Jain's composite fermion approach in Eq. \ref{wfcomposite fermion}, which is constructed from the filled  LLL and 1LL,  is exactly equal to that given by the CFT correlator in Eq. \ref{J01} up to a constant. Since the LLL single-particle wave function has no $ \bar{z} $ and the 1LL single-particle wave function  has  $ \bar{z} $ to the power of 1, which is just  $2\partial_z $ in the process of LLL projection \cite{jainbook}, we can attribute the vertex operator $ V_0 $ to LLL and  $ V_1 $ to 1LL since $ V_0 $ contains no derivative and    the power of the derivative in $ V_1$ is 1.

\section{General construction of vertex operator for composite fermions}\label{gcft} As we stated in Sec. \ref{jaincf}, Jain has chosen two filled composite fermion LLs as LLL and 1LL. This raises an interesting question: can we choose composite fermions to fill other Landau levels and obtain a new  state  at the same filling factor  $2/5$, which can still be projected to LLL? Motivated by this question and the connection of CFT vertex operators to composite fermion LLs, we can generally introduce two species of vertex operators for the $ n_1 $th and $ n_2 $th composite fermion LLs(without loss of generality, we can let $ 0\leq n_1<n_2 $) compactified on two circles of radii $ \sqrt{3} $  and $ \sqrt{15} $, respectively, by taking into account the power of $\bar{z}$ in the single-particle wave functions of the $ n_1 $th  and $ n_2 $th LL, \be\label{des2} V_{n_1}(z)=:\partial_{z}^{n_1}e^{i\sqrt{3}\phi_1(z)}:\ee and \be\label{des3} V_{n_2}(z)=:\partial_{z}^{n_2}\left( e^{i\sqrt{\frac{4}{3}}\phi_1(z)}e^{i\sqrt{\frac{5}{3}}\phi_2(z)}\right):.\ee
It is easy to demonstrate that they still represent two independent fermionic operators and their individual electric charge is still 1.
\begin{widetext}Therefore we can construct  a conformal correlator,\be \mathcal{A}\{\langle V_{n_2}(z_1)V_{n_2}(z_2)\cdots V_{n_2}(z_{\frac{N}{2}})V_{n_1}(z_{\frac{N}{2}+1})V_{n_1}(z_{\frac{N}{2}+2})\cdots V_{n_1}(z_{N})\rangle \}, \ee which is simplified as \be\label{CFT1}\mathcal{A}\{\partial_{z_1}^{n_2}\partial_{z_2}^{n_2}\cdots \partial_{z_{  \frac{N}{2} }}^{n_2}\partial_{z_{  \frac{N}{2}+1 }}^{n_1}\partial_{z_{  \frac{N}{2}+2 }}^{n_1}\cdots \partial_{z_{N }}^{n_1}\prod\limits_{i<j\leq   \frac{N}{2}   }(z_i-z_j)^3 \prod\limits_{ \frac{N}{2} <k<l }(z_k-z_l)^3 \prod\limits_{m\leq  \frac{N}{2}   <n }(z_m-z_n)^2    \}.\ee \end{widetext}

\begin{figure}\includegraphics[trim=0cm 6cm 0cm 6cm, clip=true, scale=0.35,]{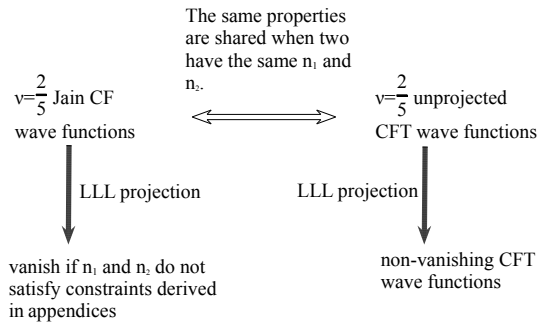}\centering
\caption{The relation between wave functions obtained from Jain's composite fermion(CF) theory and conformal field theory(CFT).}\label{JaCFT}
\end{figure}

An interesting question can be raised as to whether the state in Eq. \ref{CFT1} of general choices of $ n_1 $ and $ n_2 $ is  equal to the LLL-projected Jain composite fermion state constructed from the filled $ n_1 $th and $ n_2  $th LLs. The answer is negative unless $ n_1=0,n_2=1 $.
Explicitly,  the LLL-projected Jain composite fermion state on a disk constructed from  the filled    $ n_1 $th and $ n_2  $th LLs has the wave function \be\label{CF25Jain}\begin{split} & P_{\text{LLL}} \prod\limits_{1\leq i<j\leq N}(z_i-z_j)^2\\& \times\left| {\begin{array}{*{20}{c}}
  \eta_{n_1,-n_1}(z_1)& \ldots &\eta_{n_1,-n_1}(z_N) \\
   \vdots & \ddots & \vdots  \\
 \eta_{n_1,\frac{N}{2}-n_1-1}(z_1)& \ldots &\eta_{n_1,\frac{N}{2}-n_1-1}(z_N)\\\eta_{n_2,-n_2}(z_1)& \ldots &\eta_{n_2, -n_2}(z_N) \\
    \vdots & \ddots & \vdots  \\
 \eta_{n_2,\frac{N}{2}-n_2-1}(z_1)& \ldots &\eta_{n_2,\frac{N}{2}-n_2-1}(z_N)
\end{array}} \right|.\end{split}\ee  as in Eq. \ref{25disk}. On the other hand, the wave function constructed from CFT  in Eq. \ref{CFT1} can be shown to be equal to  \be\label{CF25CFT} P_{\text{LLL}} \Psi_{\text{CFT}}, \ee
where
\be\label{unprojCFT} \Psi_{\text{CFT}}=\prod\limits_{1\leq i<j\leq N}(z_i-z_j)^2 \left| {\begin{array}{*{20}{c}}
 \bar z_1^{n_1}& \ldots &\bar z_N^{n_1} \\
   \vdots & \ddots & \vdots  \\
 \bar z_1^{n_1}z_1^{\frac{N}{2}-1}& \ldots &\bar z_N^{n_1}z_N^{\frac{N}{2}-1}\\\bar z_1^{n_2} & \ldots &\bar z_N^{n_2}  \\
    \vdots & \ddots & \vdots  \\
\bar z_1^{n_2}z_1^{\frac{N}{2}-1}& \ldots &\bar z_N^{n_2}z_1^{\frac{N}{2}-1}
\end{array}} \right|\ee up to a constant. We term this $\Psi_{\text{CFT}}$ the unprojected CFT wave functions.
The only difference between the two wave functions in Eqs.\ref{CF25Jain}, \ref{CF25CFT} is that in contrast to the former, the latter only keeps $ \bar z^n z^{n+m} $ in each $ \eta_{n,m}(z)$, where we omit the Gaussian factor for simplicity. In Appendix  \ref{LLP}, we have proved that there are  some choices of $ n_1 $ and $ n_2 $ for which   Jain's $ \nu=2/5  $ composite fermion state has vanishing LLL projection if the constraints in Eq. \ref{nonvan} are not satisfied.
By contrast, the CFT wave function in Eq. \ref{CFT1} is   nonvanishing for general $ n_1 $ and $ n_2 $. Therefore, although some of the LLL-projected composite fermion states in Eq. \ref{CF25Jain} vanish, we can still construct nonvanishing wave functions by CFT from the same filled composite fermion LLs $ n_1 $ and $ n_2 $. Moreover, it is easy to see that before LLL projection, the unprojected Jain composite fermion state in Eq. \ref{CF25Jain} has the same root patterns as those of the corresponding unprojected CFT state in Eq. \ref{unprojCFT}(see Appendix \ref{RP} for the definition of root pattern). This strongly suggests that they have the same topological feature. As will be seen in Sec. \ref{shf},   they   have the same topological shift if they have the same $n_1$ and $n_2$, since the topological shift is dictated by the root pattern. We thus term the wave function given by CFT in Eq. \ref{CFT1} as Jain $n_1n_2$ wave function. We can  safely do so since wave functions constructed from CFT correlators are described by the same Chern-Simons field theory which characterizes the $ \nu=2/5 $ composite fermion states. The justification for this follows Ref. \onlinecite{CFTJain2}: From the CFT Lagrangian characterizing wave functions in Eq. \ref{CFT1}, we can change the basis from  massless free boson fields $\phi_1$ and  $\phi_2  $ to  $ \chi_1 =\phi_2 \sqrt{3/5}$ and  $\chi_2 =\phi_1 / \sqrt{3} - \phi_2 \sqrt{4/15}$ such that the two independent quasihole operators of charge $ 1/5 $ are $ e^{i\chi_1}$ and  $e^{i\chi_2} $. It is easy to check that $[Q,e^{i\chi_1}]=\frac{1}{5}e^{i\chi_1}$ and $[Q,e^{i\chi_2}]=\frac{1}{5}e^{i\chi_2}$. After this change of basis, we arrive at the Chern-Simons Lagrangian characterizing the $ \nu=2/5 $ state  with $ K $ matrix  $ \left( \begin{smallmatrix} 3 & 2 \\ 2 & 3 \end{smallmatrix} \right )$ and charge vector $t=(1,1)$. Note that Jain's $ 2/5 $ composite fermion states in Chern-Simons field theory, \emph{regardless of which two composite fermion LLs are filled}, are characterized by the same  $ K $ matrix $ \left( \begin{smallmatrix} 3 & 2 \\ 2 & 3 \end{smallmatrix} \right )$ and charge vector $(1,1)$ \cite{zee}. Now we can safely  argue that although the  LLL-projected Jain states might vanish, we still have    wave functions completely in LLL  constructed from CFT,  which capture the same important features such as filling factor $2/5$, fractional charge $1/5$ for quasiholes and Abelian exchange statistics for the quasihole manifold of corresponding Jain states since these features are dictated by Chern-Simons field theory. In fact, by using Eq. 6.65 in Ref. \onlinecite{Francesco_conformal}, we find that exchanging two quasiholes both of type $ e^{i\chi_1}$ or both of type $ e^{i\chi_2}$  induces a phase factor  $ e^{i3\pi /5}$ while exchanging one quasihole of type $ e^{i\chi_1}$ and another one of  type $ e^{i\chi_2}$ induces a phase factor  $ e^{-i2\pi /5}$.  We summarize the relation between the wave functions obtained from these two approaches in Fig. \ref{JaCFT}.

\section{Topological  Shift}\label{shf}
We have obtained a series of Jain $n_1n_2$  wave functions in Eq. \ref{CFT1}. Now the question is how to distinguish one from another among them. We will resort to a number called the topological shift which can differentiate one topological state from another. We can place the quantum Hall system on any 2D surface such as a disk or the surface of a sphere.
On the surface of a sphere, the number of magnetic flux quanta piercing the sphere $ N_{\Phi} $ is related to the electron number $ N_e $ by the following identity: \be  N_{\Phi}=\frac{1}{\nu} N_e-\mathcal{S},\ee where $ \nu $ is the filling factor and $ \mathcal{S} $ is the topological shift characterizing the state.

The topological shift has a measurable effect: it is related to the Hall viscosity $ \eta^{(A)} $ by the relation, \cite{vis,RR11}\be\eta^{(A)}=\frac{1}{4}\mathcal{S} n\hbar, \ee where $ n $ is the average electron density. The topological shift can be verified by experiments in which the Hall viscosity is measured. Recently experimental protocols for such measurements have been proposed \cite{H_vis,H_vis2,H_vis3,98.161303,2019arXiv190503269M}. Also noteworthy is the finding that interfaces separating states with different Hall viscosities carry electric dipole moments, which can have important physics consequences \cite{Zhu_arXiv200302621Z}.

Topological shift can be easily  read off from the root pattern of wave function of the state on the sphere. For example, consider the case in which  only the $ n $th LL is  completely filled, where $ n $ can be any non-negative integer. Its root pattern is $1_n 1_n 1_n ...1_n$, where 1 denotes an occupied orbital. The angular momentum of the $ n $th LL on the sphere is $ N_{\Phi}/2+n $. Hence the minimum and maximum angular momenta of occupied orbitals in the root pattern are $ -(N_{\Phi}/2+n) $ and $ (N_{\Phi}/2+n) $, respectively. Therefore, we have \be N_e= 2(N_{\Phi}/2+n)+1.\ee We thus get the shift $ 2n+1 $ for  the completely filled $ n $th LL with $ n=0,1,2... $.

Another example is  the 1/3 Laughlin   state in LLL, whose root pattern is 100100100...1, in which 0 stands for the unoccupied orbital and we have neglected LL indices for simplicity. The minimum and maximum angular momenta in LLL on the sphere are $ -N_{\Phi}/2 $ and $ N_{\Phi}/2 $, respectively.  We have \be N_{\Phi}/2-(-N_{\Phi}/2)=3(N_e-1).\ee  Thus the topological shift is 3 for the 1/3 Laughlin state in LLL.

For states constructed from CFT, the boundary terms in their root patterns are more and more complicated as $n_1$ and $n_2$ increase. Their shifts are derived in the following way. Since Jain composite fermion state constructed from Landau levels  $n_1$ and $n_2$ has the same root pattern as that of the corresponding CFT state(This fact can be easily deduced from Eq. \ref{CF25Jain} and  Eq. \ref{CF25CFT}), the topological shift is the same for both states as long as they have the same $n_1$ and $n_2$.  So we can use the explicit form of Jain's composite fermion wave function to calculate the  topological shift for the state constructed from CFT.
It is known that the monopole charge $Q$ of the $\nu=2/5$ composite fermion state is related to the monopole charge $Q^*$ of the $\nu=2 $ integer quantum Hall state by the identity \cite{jainbook} $Q=Q^*+N-1$ where $N$ is the particle number(see also Eq. \ref{Q}). Since we have the  $n_1$th and $n_2$th LLs filled, the particle number is \be N=2(Q^*+n_1)+1+2(Q^*+n_2)+1.\ee
The number of magnetic flux quanta is \be N_{\Phi}=2Q=2(Q^*+N-1)=\frac{5}{2}N-(n_1+n_2+3).\ee
Hence the topological shift is $n_1+n_2+3$ for the  $\nu=2/5$ Jain $n_1n_2$ state constructed from CFT.  Now a confusion arises that two Jain $n_1n_2$ states of the same $n_1+n_2$  have the same topological shift. A single topological number such as shift is insufficient to differentiate two topologically distinct states; we thus must resort to other topological numbers as well.

\section{multibody relative angular momentum}\label{mb}
Quantum Hall states are also characterized by a set of numbers $\{S_n\} $ named the pattern of zeros(PZ) \cite{WenWang1,WenWang2}. These $S_n$ are the minimum $n$-body relative angular momentum(unit of angular momentum is chosen as $\hbar$) of a specific quantum Hall state. To obtain  $S_n$, we let $z_i$ with $i=2,...n$ approach $z_1$ in the polynomial part of the wave function  $\Psi(z_1,z_2,\dotsc,z_N)$ of  the concerned quantum Hall state  and then collect the power of  $\lambda$ in the leading term, \be\begin{split}
& \Psi(z_1,z_2=z_1+\lambda \eta_2,\dotsc, z_n=z_1+\lambda \eta_n, z_{n+1},\dotsc,z_N)\\ =& \lambda^{S_n}f(z_1,\eta_2,\dotsc, \eta_n,z_{n+1},\dotsc,z_N)+\mathcal {O}(\lambda^{S_n+1}).
\end{split} \ee $S_n$  of various Jain states are tabulated in Table \ref{table:poz}.
\begin{table*}[hbt]
\begin{tabular}{ m{1.5cm}  m{1.0cm}  m{1.0cm}  m{1.0cm} m{1.0cm} m{1.0cm} m{1.0cm} m{1.0cm} m{1.0cm}}
\hline\hline
  &  $S_2$   &   $S_3$  &  $S_4$   & $S_5$ & $S_6$& $S_7$& $S_8$ &...\\ [0.5ex] 
\hline\\
Jain 01 & 1  & 5 & 12 & 21 & 33 & 47 & 64\\[0.5ex]
Jain 02 & 1  & 3 & 9 & 18 & 29& 43 & 59\\[0.5ex]
Jain 12 & 1  & 3 & 8 & 16 & 27& 40 & 56\\[0.5ex]
Jain 03 & 1  & 3 & 6 & 14 & 25& 38 & 54\\[0.5ex]
Jain 13 & 1  & 3 & 6 & 13 & 23& 36 & 51\\[0.5ex]
\hline
\end{tabular}
\caption{The pattern of zeros of various Jain states. In constructing these states, we let the number of composite fermions  in each of two composite fermion LLs be equal. Here we list $S_n$ from $n=2$ to 8 as these are sufficient to distinguish one state from another.}\label{table:poz}
\end{table*}


From PZ we may  obtain projectors for which a specific Jain  $n_1n_2$ state is a zero-energy ground state, although not the unique zero-energy ground state in its own sector. For example, the minimum three-body relative angular momentum $S_3$ of the Jain 01 state is 5, so the Jain 01 state is a zero-energy ground state of a three-body operator $P^{(3)}_3$ which projects on the three-body antisymmetric state of relative angular momentum 3. \footnote{Due to  antisymmetry of fermionic wave functions, the  operator $P^{(4)}_3$ which projects on the three-body antisymmetric state of relative angular momentum 4 does not exist}   However, the Jain 01 state is not the unique zero-energy ground state of  $P^{(3)}_3$  in its own sector.  In fact, the  densest zero-energy ground state of $P^{(3)}_3$ is the Pfaffian state \cite{MR,GreiterPf} with filling factor $\nu=1/2$, which has root pattern 110011001100.... The minimum four-body relative angular momentum $S_4$ of the Jain 01 state is 12, so the Jain 01 state is also a zero-energy ground state of   four-body projection operators $P^{(6)}_4$,  $P^{(8)}_4$,  $P^{(9)}_4$,  $P^{(10)}_{4}$ and $P^{(11)}_{4}$  which project on four-body antisymmetric states of relative angular momentum 6, 8, 9, 10 and 11, respectively. \footnote{Their explicit forms are given in Appendix \ref{projop}. Note that there are two independent $P^{(10)}_{4}$.  Reader can refer to Ref. \onlinecite{Simonetal2} for details. Again, due to  antisymmetry of fermionic wave functions, the  operator $P^{(7)}_4$  does not exist} If we choose the Hamiltonian as a linear combination of $P^{(3)}_3$,  $P^{(10)}_{4}$ and $P^{(11)}_{4}$ with positive coefficients(this results from the fact that Pfaffian state, being the densest zero-energy ground state of $P^{(3)}_3$, has minimum four-body relative angular momentum 10; thus the Pfaffian state is automatically annihilated by $P^{(6)}_4$,  $P^{(8)}_4$ and $P^{(9)}_4$), 
again the Jain 01 state  is not the unique zero-energy ground state in its sector. 
As a by-product, we have found the densest zero-energy ground state of the above Hamiltonian to have filling factor $\nu=3/7$, which is larger than $2/5$. Detailed discussion on this $\nu=3/7$ state will be given in the next section.

Similarly, the Jain 02 state is a zero-energy ground state of $P^{(6)}_4$ and  $P^{(8)}_4$, although not the unique zero-energy ground state in its sector. We have chosen a linear combination of $P^{(6)}_4$ and  $P^{(8)}_4$ with positive coefficients as our parent Hamiltonian and diagonalized  it for up to 10 particles on the disk. 
The densest zero-energy ground state of this Hamiltonian is found to be denser than the Jain 02 state. We will also discuss this in the next section.

Likewise, the Jain 12 state is a zero-energy ground state of $P^{(6)}_4$, although not the unique zero-energy ground state in its sector. In fact, the unique densest zero-energy ground state of $P^{(6)}_4$ is the Read-Rezayi $\mathbb{Z}_3$ state \cite{RR} with  $\nu=3/5$.

Similar arguments can be made about other Jain $n_1n_2$ states as well. We conjecture that it is impossible to find parent Hamiltonians for which a Jain $n_1n_2$ state is the densest zero-energy ground state. This has been discussed in Refs. \onlinecite{Chen17,exactH} for the Jain $01$ state, which is exactly equal to the LLL-projected Jain CF state constructed from CF 0LL and 1LL.
In Ref. \onlinecite{Chen17}, it is argued that an exact parent Hamiltonian which can give correct edge mode counting for the Jain $01$ state is nonexistent due to the reduced degree of freedom in Landau levels after LLL projection. It has been conjectured in Ref. \onlinecite{exactH} that the  impossibility of finding the exact parent Hamiltonian could be  related to descendant fields in the CFT correlator. Indeed, introducing descendant fields  as in Eqs. \ref{des2}, \ref{des3}  consequently introduces derivatives to the wave function, thus  the wave function does not have property of heredity when the particle number of the system increases by 1. Below we introduce the notion of heredity. In our previous work on  the Laughlin state and unprojected Jain $\nu=2/5$ state \cite{Chen14, Chen19}, each of which is the unique densest zero-energy ground state of a parent Hamiltonian, it is found that  recursive formula  in particle number $N$ for the densest zero-energy ground state of the parent Hamiltonian  is of  the following  form, \be \ket{\psi_{N+1}} \propto \sum_{m} c_{n,m}^\dag  G_{r_{\text{max}}-m}\ket{\psi_{N}}.\ee Here $r_{\text{max}} $ is the maximum occupied orbital in $\ket{\psi_{N+1}}$ and $G_{r_{\text{max}}-m}$ is some zero mode  generator which gives a new zero mode when acting on an existing zero mode if $r_{\text{max}}>m$. If $r_{\text{max}}=m$, $G_0$ is defined as the identity operator. $G_{r_{\text{max}}-m}$ is automatically set to 0 if $r_{\text{max}}<m$.
Therefore, $\ket{\psi_{N+1}}$ contains terms  proportional to $c_{n,r_{\text{max}}}^\dag   \ket{\psi_{N}}$. We term this property  of the wave function the heredity when the particle number of the system increases.  Due to the existence of derivatives in their wave function, all Jain $n_1n_2$ states do not possess the property of heredity. Using the method of contradiction,   the parent Hamiltonians for these states do not exist. Otherwise, they would have a recursive formula   following the same logic as in Refs. \onlinecite{Chen14,Chen19} and thus would have the property of heredity.

\section{The densest zero-energy ground states of certain projection operators}\label{thrsvn}
In the previous section, we have searched for the parent Hamiltonians of various Jain $n_1n_2$ states, based on their minimum multibody relative angular momenta or PZ. We find that although general Jain $n_1n_2$ states are zero-energy ground states of certain projection operators, they are not the densest ones. In studying parent Hamiltonians in the form of projectors, we have found two interesting states which are worth discussing here.

\begin{figure}\includegraphics[trim=-3cm 3cm -14cm 3cm, clip=true, scale=0.35,]{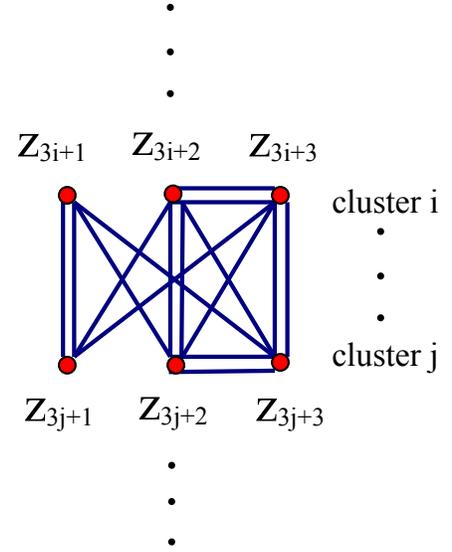}\centering
\caption{The pairings of the intracluster and intercluster parts of the wave function $\psi_b$ in Eq. \ref{threeseven} for    clusters $i$ and $j$. Red dots denote particles and black dots represent clusters other than $i$ and $j$. Particles connected by one  line contribute their relative coordinate to the power of 1 to the wave function  while particles connected by two lines contribute their relative coordinate to the power of 2 to the wave function. }\label{clus}
\end{figure}

We haven chosen the Hamiltonian as a linear combination of $P^{(3)}_3$,  $P^{(10)}_{4}$ and $P^{(11)}_{4}$ with positive coefficients and diagonalized it for up to 7 particles on the disk and found one densest zero-energy ground state at particle number 5 and 6. The root pattern of this state is 11001001100100..., which has repetitions of 1100100. For  particle number 4 and 7, there is an extra independent zero-energy ground state in each case, with root pattern    1100011 and 11001001100011, respectively.  For the above mentioned $\nu=3/7$ state with root pattern 11001001100100..., we have found its wave function on the disk to have the following form(Gaussian factor omitted), \be\label{threeseven} \psi_{\nu=\frac{3}{7}}=\psi_b \prod\limits_{i<j}(z_i-z_j),\ee where $ \psi_b $ is a bosonic wave function in which every three particles form a cluster. Its explicit form is given in the following way. First we divide particles into clusters, with each cluster having three particles.  We then choose any two clusters,   whose particle coordinates are $z_{3i+1}, z_{3i+2}, z_{3i+3}$ and  $z_{3j+1}, z_{3j+2}, z_{3j+3}$, respectively. We  assign to these two clusters an intercluster wave function \be\begin{split}
&(z_{3i+1}-z_{3j+1})^2(z_{3i+1}-z_{3j+2})(z_{3i+1}-z_{3j+3})\\&(z_{3i+2}-z_{3j+1})(z_{3i+2}-z_{3j+2})^2(z_{3i+2}-z_{3j+3})\\&(z_{3i+3}-z_{3j+1})(z_{3i+3}-z_{3j+2})(z_{3i+3}-z_{3j+3})^2.
\end{split}\ee To the cluster   of particle coordinates   $z_{3i+1}, z_{3i+2}$ and $z_{3i+3}$, we assign an intracluster wave function $(z_{3i+2}-z_{3i+3})^2$. Finally, we symmetrize the product of all intercluster    and intracluster wave functions to obtain the bosonic wave function $\psi_b$.   The pairings of the intracluster and intercluster parts of the wave function for  two general clusters are shown in Fig. \ref{clus}. It is easy to verify that $\psi_{\nu=\frac{3}{7}} $ has  minimum $3$-body relative angular momentum $S_3=5$  and minimum $4$-body relative angular momentum $S_4=12$; thus it is indeed a zero-energy ground state of $P^{(3)}_3$,  $P^{(10)}_{4}$ and $P^{(11)}_{4}$.  We note that this 3/7 state is distinct from the $S_3$ state \cite{S3} and the Jain state at the same filling factor $\nu=3/7$: both the $S_3$ state and the Jain state have integer topological shift while this 3/7 state has fractional topological shift. \footnote{It turns out this 3/7 state had been discovered independently by Thomale et al. \cite{TBG,Bo241302,Bo1907}.}

We have also chosen a linear combination of $P^{(6)}_4$ and  $P^{(8)}_4$ with positive coefficients as our parent Hamiltonian and diagonalized  it for up to 10 particles on disk. We find a unique densest zero-energy ground state for 4, 5, 6, 8 and 10 particles, respectively.\footnote{See Supplemental Material at [URL will be inserted by publisher] for its second-quantized form.} For 7 particles, there are two densest zero-energy ground states with the same angular momentum. For 9 particles, there are three  densest zero-energy ground states with the same angular momentum. 
The unique densest state has root pattern 111000110100110011, whose filling factor is very close to $1/2$ for  a finite number of particles as seen from  the root pattern. This state has minimum $4$-body relative angular momentum $S_4=9$; thus it is indeed a zero-energy ground state of $P^{(6)}_4$ and  $P^{(8)}_4$.
Note that the  Pfaffian state  is a candidate for the zero-energy ground state of this parent Hamiltonian \cite{Simonetal}. If the  state with root pattern 111000110100110011 remains gapped  and possesses $\nu=1/2$  in the thermodynamic limit, it would be  of interest to study its properties, obtain the close form for its first-quantized wave function,  and compare it with other $\nu=1/2$   states such as Pfaffian, anti-Pfaffian \cite{levinAP,leeAP} and PH-Pfaffian \cite{Son, Zucker,Feldmanhalfinteger}.

\section{Conclusion and Discussion}

In this paper, we have constructed a series of $\nu=2/5$ fractional quantum Hall trial wave functions from CFT, and discovered their one-to-one correspondence with Jain's composite fermion wave functions constructed using different composite fermion Landau levels (LLs). The forms of CFT wave functions are simpler than those of composite fermion wave functions as seen in Eqs. \ref{CF25Jain} and \ref{unprojCFT}, yet the former and the latter have the same topological properties as long as they are constructed from the same two  composite fermion LLs.
Among $\nu=2/5$ CFT wave functions, those  corresponding to  different composite fermion LLs  are characterized by topological shifts and multibody relative angular momenta. One thing we need to pay special attention to is that  filling factors of all these   CFT wave functions  are not exactly  $ 2/5$ for  a finite number of particles. In fact, their filling factors deviate from $ 2/5$ for  a finite number of particles and approach $ 2/5$ only in the  thermodynamic limit. Furthermore, their exact forms are  difficult to deal with  for large numbers of particles due to the action of taking derivatives and subsequent antisymmetrization. As a result, it is not easy to calculate their excitation energies over the ground state in their individual sector.  A possible  way to circumvent this difficulty is to approximate CFT wave functions from Eq. \ref{unprojCFT} via Jain's approximate projection \cite{JK1,JK2}. Thus we leave the task of calculating  their energies over the Coulomb ground state to future work.

In the process of trying to find parent Hamiltonians  for these CFT wave functions, we have discovered  two interesting states. One is a $\nu=3/7$ state as the densest zero-energy ground state of $P^{(3)}_3$,  $P^{(10)}_{4}$ and $P^{(11)}_{4}$. Its first-quantized wave function given in Eq. \ref{threeseven} involves clusters of three particles, but the way it goes to zero when several particles come together is obviously distinct from that for the Read-Rezayi $\mathbb{Z}_3$ state. It will be of interest to study its CFT nature in the future.
The other one is  the unique densest zero-energy ground state of  $P^{(6)}_4$ and  $P^{(8)}_4$, for which we only get the second-quantized wave function for up to 10 particles. This state has root pattern 111000110100110011, with a filling factor close to $1/2$ for  a finite number of particles. It is worth studying whether its filling factor is  $1/2$ in the thermodynamic limit. If so, it would be useful to study its gap and compare this state with other $\nu=1/2$ candidate states.

\begin{acknowledgments}
Part of this work is supported by US DOE, Office of BES through Grant No. DE-SC0002140, and performed at the National High Magnetic Field Laboratory, which is supported by NSF Cooperative Agreements No. DMR-1157490 and DMR-1644779, and the State of Florida. L.C. acknowledges support from NSFC Grant No. 11947027. The authors also thank Alexander Seidel and Duncan Haldane for helpful discussions.\end{acknowledgments}

\appendix
\section{Root pattern}\label{RP} We can always expand many-body quantum Hall wave functions in terms of Slater determinants of simultaneous  single-particle eigenstates of a one-body Hamiltonian and a one-body operator reflecting the symmetry of the geometry in which the quantum Hall system resides(for example, on the disk this operator is the single-particle angular momentum operator),
\be
\ket {\psi} = \sum_{\{n\}}C_{\{n\}} |\{n\}\rangle\,.
\ee
where $\ket{\{n\}}$ is such a Slater determinant. This Slater determinant is labeled by patterns such as $1_{n_0}01_{n_1}1_{n_2}0...$, where 1 denotes an occupied orbital and 0 denotes an unoccupied orbital. The subscripts $n_0, n_1...$ are LL indices of occupied orbitals. Of all patterns resulting from such a Slater determinant expansion, there is  a special one known as the root pattern in the sense that all patterns other than the root pattern can be obtained from the root pattern via  inward squeezing \cite{ortiz,Chen17}.
Inward squeezing involves  inward pair hoppings of two particles while conserving center of mass of orbitals of these two particles. For example, $01_{n_1}1_{n_2}0$ can be obtained from $1_{n_3}001_{n_4}$ via inward squeezing. Note that inward squeezing can change  LL indices of occupied orbitals.
Another example is the Slater determinant expansion of the $\nu=1/3$ Laughlin state of 3 particles \cite{slater} on the  disk. Since this state resides entirely in LLL,  we omit LL indices below for simplicity. Patterns of Slater determinants in the expansion are $1001001$, $0110001$, $1000110$, $0101010$, and   $0011100$. Here $1001001$ is the root pattern while all other patterns can be obtained from it via inward squeezing.

\section{Three-body and four-body relative angular momentum projectors on disk}\label{projop}
The first-quantized  three-particle  states of general  relative angular momentum are  given in Ref. \onlinecite{Simonetal2}.
Here we give the explicit second-quantized form of operators which projects onto a three-body state of relative angular momentum 3 in LLL  on the disk,\be P^{(3)}_3=\sum\limits_{3R-3 \in \mathbb{N}} Q^{(3)\dag}_R  Q^{(3)}_R,\ee
with
\be\begin{split}
Q^{(3)}_R=\sum_{\substack{i_1<i_2<i_3\\ i_1+i_2+i_3=3R}}(i_1-i_2) (i_1-i_3) (i_2-i_3)\\\times\sqrt{\frac{(3R-3)!}{i_1!i_2!i_3!}}c_{0,i_3}c_{0,i_2}c_{0,i_1}.
\end{split} \ee
$R$ is the center-of-mass angular momentum and  $ c_{0,i} $ is a fermionic operator which annihilates a fermion of angular momentum $ i $ in LLL.
Following the way in Ref. \onlinecite{Simonetal2}, we can also easily derive the  second-quantized form of operators, each of which projects onto the four-body state in LLL  on a disk of relative angular momentum 6, 8, 9,10 and 11, respectively, \begin{widetext}\be
P^{(6)}_4=\sum\limits_{4R-6 \in \mathbb{N}} S^{(6)\dag}_R  S^{(6)}_R,\ee
with
\be\begin{split} S^{(6)}_R=\sum_{\substack{i_1<i_2<i_3<i_4\\ i_1+i_2+i_3+i_4=4R}}&(i_1-i_2) (i_1-i_3)(i_1-i_4) (i_2-i_3)(i_2-i_4)(i_3-i_4) \sqrt{\frac{(4R-6)!}{i_1!i_2!i_3!i_4!}}c_{0,i_4}c_{0,i_3}c_{0,i_2}c_{0,i_1},\end{split}\ee
\be P^{(8)}_4=\sum\limits_{4R-8 \in \mathbb{N}} S^{(8)\dag}_R  S^{(8)}_R,\ee
with
\be\begin{split} S^{(8)}_R=\sum_{\substack{i_1<i_2<i_3<i_4\\ i_1+i_2+i_3+i_4=4R}} & f_{8}(i_1,i_2,i_3,i_4)(i_1-i_2) (i_1-i_3)(i_1-i_4) (i_2-i_3)(i_2-i_4)(i_3-i_4)\\&\times \sqrt{\frac{(4R-8)!}{i_1!i_2!i_3!i_4!}}c_{0,i_4}c_{0,i_3}c_{0,i_2}c_{0,i_1},\end{split}\ee

\be P^{(9)}_4=\sum\limits_{4R-9 \in \mathbb{N}} S^{(9)\dag}_R  S^{(9)}_R,\ee
with
\be\begin{split} S^{(9)}_R=\sum_{\substack{i_1<i_2<i_3<i_4\\ i_1+i_2+i_3+i_4=4R}} & f_{9}(i_1,i_2,i_3,i_4)(i_1-i_2) (i_1-i_3)(i_1-i_4) (i_2-i_3)(i_2-i_4)(i_3-i_4)\\&\times \sqrt{\frac{(4R-9)!}{i_1!i_2!i_3!i_4!}}c_{0,i_4}c_{0,i_3}c_{0,i_2}c_{0,i_1},\end{split}\ee

\be P^{(10)}_4=\sum\limits_{4R-10 \in \mathbb{N}} S^{(10)\dag}_R  S^{(10)}_R,\ee
with
\be\begin{split} S^{(10)}_R=\sum_{\substack{i_1<i_2<i_3<i_4\\ i_1+i_2+i_3+i_4=4R}} & f_{10}(i_1,i_2,i_3,i_4)(i_1-i_2) (i_1-i_3)(i_1-i_4) (i_2-i_3)(i_2-i_4)(i_3-i_4)\\&\times \sqrt{\frac{(4R-10)!}{i_1!i_2!i_3!i_4!}}c_{0,i_4}c_{0,i_3}c_{0,i_2}c_{0,i_1},\end{split}\ee

\be P^{(10)'}_4=\sum\limits_{4R-10 \in \mathbb{N}} S^{(10)'\dag}_R  S^{(10)'}_R,\ee
with
\be\begin{split} S^{(10)'}_R=\sum_{\substack{i_1<i_2<i_3<i_4\\ i_1+i_2+i_3+i_4=4R}} & f'_{10}(i_1,i_2,i_3,i_4)(i_1-i_2) (i_1-i_3)(i_1-i_4) (i_2-i_3)(i_2-i_4)(i_3-i_4)\\&\times \sqrt{\frac{(4R-10)!}{i_1!i_2!i_3!i_4!}}c_{0,i_4}c_{0,i_3}c_{0,i_2}c_{0,i_1},\end{split}\ee
and \be P^{(11)}_4=\sum\limits_{4R-11 \in \mathbb{N}} S^{(11)\dag}_R  S^{(11)}_R,\ee
with
\be\begin{split} S^{(11)}_R=\sum_{\substack{i_1<i_2<i_3<i_4\\ i_1+i_2+i_3+i_4=4R}} & f_{11}(i_1,i_2,i_3,i_4)(i_1-i_2) (i_1-i_3)(i_1-i_4) (i_2-i_3)(i_2-i_4)(i_3-i_4)\\&\times \sqrt{\frac{(4R-11)!}{i_1!i_2!i_3!i_4!}}c_{0,i_4}c_{0,i_3}c_{0,i_2}c_{0,i_1}.\end{split}\ee \end{widetext} Coefficients $f_{8}(i_1,i_2,i_3,i_4)$, $f_{9}(i_1,i_2,i_3,i_4)$, $f_{10}(i_1,i_2,i_3,i_4)$, $f'_{10}(i_1,i_2,i_3,i_4)$,   and   $f_{11}(i_1,i_2,i_3,i_4)$  can be expressed in terms of elementary symmetric polynomials in $i_1$, $i_2$, $i_3$, and $i_4$:

\begin{subequations}
\be
\begin{split}
f_{8}=3 e_1^2-8 e_2-15 e_1+70,
\end{split}
\ee\be
\begin{split}
f_{9}=& e_1^3-4 e_2 e_1+8 e_3 -9 e_1^2+24 e_2+30 e_1 -120,
\end{split}
\ee
\be
\begin{split}
f_{10}=& 3 e_1^4-16 e_2 e_1^2+20 e_2^2+4 e_3 e_1-16 e_4\\& -42 e_1^3+128 e_2 e_1-96 e_3\\& +321 e_1^2-656 e_2-1030 e_1+2924,
\end{split}
\ee
\be
\begin{split}
f'_{10}=& 3 e_1^4-16 e_2 e_1^2+16 e_2^2+16 e_3 e_1-64 e_4 \\& -42 e_1^3+144 e_2 e_1 -192 e_3\\& +285 e_1^2-640 e_2-870 e_1+2656,
\end{split}
\ee\be
\begin{split}
f_{11}=& 3 e_1^5-20 e_2 e_1^3+24 e_3 e_1^2+32 e_2^2 e_1-64 e_2 e_3\\& -60 e_1^4+292 e_2 e_1^2-256 e_2^2-232 e_3 e_1+256 e_4\\& +553 e_1^3-1792 e_2 e_1+1904 e_3\\& -2940 e_1^2+6160 e_2 +7980 e_1-20944,
\end{split}
\ee
\end{subequations} where four elementary symmetric polynomials in $i_1$, $i_2$, $i_3$, and $i_4$ are  $e_1=i_1+i_2+i_3+i_4$, $e_2=i_1 i_2+i_1 i_3+i_1 i_4+i_2 i_3+i_2 i_4+i_3 i_4$, $e_3=i_2 i_3 i_4+i_1 i_3 i_4+i_1 i_2 i_4+i_1 i_2 i_3 $,  and $e_4=i_1 i_2 i_3 i_4$.

Note that there are two independent four-particle fermionic  states of  relative angular momentum 10.
In the above equations, we have omitted the normalization factors of all $Q_R$, which only depend on center-of-mass angular momentum $R$, since we are only interested in finding the common  densest  zero-energy ground state of $Q_R$ of all possible $R$.
\section{The necessary conditions for general $ \nu=\frac{p}{2p+1} $ CF Jain states on disk  to have nonvanishing  LLL projection}\label{LLP}
Let us consider the simplest case in the first place, which is $p=2$. We begin with the $ \nu=2/5 $ Jain state constructed from    CFs   filling the $ n_1 $th and $ n_2 $th LLs. The number of CFs in the $ n_1 $th and $ n_2 $th LLs are $ N_1 $ and $ N-N_1$, respectively. $ N_1 $ must satisfy the constraint $ 1\leq N_1 \leq N-1  $. The wave function of this state is \be\label{25disk} \prod\limits_{1\leq i<j\leq N}(z_i-z_j)^2 \left| {\begin{array}{*{20}{c}}
  \eta_{n_1,m_1}(z_1)& \ldots &\eta_{n_1,m_1}(z_N) \\
   \vdots & \ddots & \vdots  \\
 \eta_{n_1,m_{N_1}}(z_1)& \ldots &\eta_{n_1,m_{N_1}}(z_N)\\\eta_{n_2,m_{N_1+1}}(z_1)& \ldots &\eta_{n_2,m_{N_1+1}}(z_N) \\
    \vdots & \ddots & \vdots  \\
 \eta_{n_2,m_{N}}(z_1)& \ldots &\eta_{n_2,m_{N}}(z_N)
\end{array}} \right|.\ee
We can expand the Laughlin-Jastrow factor $ \prod\limits_{1\leq i<j\leq N}(z_i-z_j)^2 $ as \be \sum\limits_{\substack{i_1+i_2+\cdots +i_N=(N-1)N\\ 0\leq i_1,i_2,\dotsc, i_N\leq2(N-1)}} C_{i_1,i_2,\dotsc, i_N}z_1^{i_1}z_2^{i_2}\cdots z_N^{i_N},\ee where $ C_{i_1,i_2,\dotsc, i_N} $ is the expansion coefficient.\footnote{$C_{i_1,i_2,\dotsc, i_N}$ can be obtained in two ways. For  particle number not too large, we can do exact diagonalization of a two-body  hard-core potential $\sum_{i,j} \delta(x_i-x_j)\delta(y_i-y_j) $ in LLL for which the bosonic $\nu=1/2$ Laughlin wave function $ \prod_{1\leq i<j\leq N}(z_i-z_j)^2 $ is an exact zero-energy ground state. Then from the coefficient of $b_{i_1}^\dag b_{i_2}^\dag\cdots b_{i_N}^\dag \ket{0}$ in the numerically calculated zero   energy ground state,($b_{i_1}^\dag$ is a bosonic creation operator which creates a boson of angular momentum  $i_1\hbar$ in LLL) we can obtain $C_{i_1,i_2,\dotsc, i_N}$. Alternatively, we can calculate $ C_{i_1,i_2,\dotsc, i_N} $ by using the recursive formula for bosonic $\nu=1/2$ Laughlin wave function given in Ref. \onlinecite{Chen14}}
Since  this Laughlin-Jastrow factor is symmetric in all variables, $ C_{i_1,i_2,\dotsc, i_N} $ will be invariant under the exchange of any two indices. Then the wave function can be expanded as  \be\label{ma}\begin{split}
&\sum\limits_{\substack{i_1+i_2+\cdots +i_N=(N-1)N\\0\leq i_1,i_2,\dotsc, i_N\leq2(N-1)}} C_{i_1,i_2,\dotsc, i_N}\\&\times \left| {\begin{array}{*{20}{c}}
  z_1^{i_1}\eta_{n_1,m_1}(z_1)& \ldots & z_N^{i_1}\eta_{n_1,m_1}(z_N) \\
   \vdots & \ddots & \vdots  \\
 z_1^{i_{N_1}}\eta_{n_1,m_{N_1}}(z_1)& \ldots & z_N^{i_{N_1}}\eta_{n_1,m_{N_1}}(z_N)\\z_1^{i_{N_1+1}}\eta_{n_2,m_{N_1+1}}(z_1)& \ldots & z_N^{i_{N_1+1}}\eta_{n_2,m_{N_1+1}}(z_N) \\
    \vdots & \ddots & \vdots  \\
 z_1^{i_N}\eta_{n_2,m_{N}}(z_1)& \ldots & z_N^{i_N}\eta_{n_2,m_{N}}(z_N)
\end{array}} \right|,
\end{split}\ee
where we have used this symmetry of $ C_{i_1,i_2,\dotsc, i_N} $.
Now with the identity\footnote{This identity can be easily  proved using the following formula for generalized Laguerre polynomial, $L_n^m(x)=\sum_{k=0}^{i}(-1)^k {i \choose k}L_{n-k}^{m+i}(x) $} \be\begin{split}
z^i \eta_{n,m}(z)=2^{\frac{i}{2}} \sum\limits_{k=0}^{i} &{i \choose k} \sqrt{\frac{n!(n+m+i-k)!}{(n-k)!(n+m)!}}  \\&\times\eta_{n-k,m+i}(z),\end{split} \ee
the above expansion of the wave function can be further simplified as \begin{widetext} \be\label{mat2}\begin{split}
2^{(N-1)N/2}\sum\limits_{\substack{i_1+i_2+\cdots +i_N=(N-1)N\\0\leq i_1,i_2,\dotsc, i_N\leq2(N-1)}} & C_{i_1,i_2,\dotsc, i_N}\sum\limits_{k_1=0}^{i_1}\cdots \sum\limits_{k_N=0}^{i_N}{i_1 \choose k_1} \sqrt{\frac{n_1!(n_1+m_1+i_1-k_1)!}{(n_1-k_1)!(n_1+m_1)!}}\cdots \\&\times{i_N \choose k_N} \sqrt{\frac{n_2!(n_2+m_N+i_N-k_N)!}{(n_2-k_N)!(n_2+m_N)!}} \\&\times \left| {\begin{array}{*{20}{c}}
  \eta_{n_1-k_1,m_1+i_1}(z_1)& \ldots & \eta_{n_1-k_1,m_1+i_1}(z_N) \\
   \vdots & \ddots & \vdots  \\
 \eta_{n_1-k_{N_1},m_{N_1}+i_{N_1}}(z_1)& \ldots & \eta_{n_1-k_{N_1},m_{N_1}+i_{N_1}}(z_N)\\\eta_{n_2-k_{N_1+1},m_{N_1+1}+i_{N_1+1}}(z_1)& \ldots & \eta_{n_2-k_{N_1+1},m_{N_1+1}+i_{N_1+1}}(z_N) \\
    \vdots & \ddots & \vdots  \\
 \eta_{n_2-k_N,m_{N}+i_N}(z_1)& \ldots & \eta_{n_2-k_N,m_{N}+i_N}(z_N)
\end{array}} \right|. \end{split}\ee \end{widetext} Eq. \ref{mat2} can be used to obtain composite fermion wave functions projected to any LL. We can immediately  see that in order for this wave function to  have nonvanishing LLL projection, each entry in the Slater determinant in Eq. \ref{mat2} must be an LLL single-particle wave function. We then have $k_1,k_2,\dotsc, k_{N_1}=n_1$ and $k_{N_1+1},\dotsc, k_N=n_2$. 

Thus we  must have the following constraints on $n_1$ and $n_2$,
\be \begin{split}
& n_1\leq i_1,i_2,\dotsc, i_{N_1}.\\& n_2\leq i_{N_1+1},i_{N_1+2},\dotsc, i_N.
\end{split}\ee
Since   $i_1,i_2,\dotsc, i_N$  are arbitrary, yet simultaneously satisfy two constraints, $i_1+i_2+\cdots +i_N=(N-1)N$ and $0\leq i_1,i_2,\dotsc, i_N\leq2(N-1)$, we  immediately obtain equivalent constraints, \be\label{nonvan}\begin{split} & N_1 n_1+(N-N_1 )n_2\leq (N-1)N, \quad 1\leq N_1 \leq  N-1.\\ & n_1,n_2\leq 2(N-1).\end{split} \ee


We can easily generalize this  analysis to  the $\nu= p/(2p+1) $ CF Jain state in which the $  n_1,n_2,\dotsc, n_p $th CF LLs are  filled with $  N_1,N_2,\dotsc, N_p$ CFs, respectively. The necessary conditions for this wave function to have nonvanishing LLL projection are \be\label{constr0}\begin{split}& \sum\limits_{i=1}^p N_i n_i \leq (N-1)N,\\&  n_1,n_2,\dotsc, n_p\leq 2(N-1), \end{split} \ee where $N=\sum_{i=1}^p N_i$.


We must stress that the above conditions are all \textbf{necessary}  conditions since   terms in the expansion Eq. \ref{mat2} may cancel among themselves to render vanishing LLL projection.


\section{The necessary conditions for general $ \nu=\frac{p}{2p+1} $ CF Jain states on sphere  to have nonvanishing  LLL projection}\label{LLPs}
Let us consider a  quantum Hall system on the surface of a sphere of radius $R$, subject to a radial magnetic field $\mathbf B=\frac{\hbar Q}{eR^2}\hat{\mathbf r}$, where the monopole strength $Q$ is one-half of the flux quanta number $ N_\Phi $ piercing the sphere. The  Hamiltonian of this system is \be  H=\frac{(p_x+eA_x)^2}{2m_e}+\frac{(p_y+eA_y)^2}{2m_e}+\frac{(p_z+eA_z)^2}{2m_e},\ee subject to the constraint imposed by the sphere surface  \be x^2+y^2+z^2=R^2.\ee
Using the gauge $\mathbf A= -\frac{Q}{eR}\cot\theta \mathbf{\hat\phi} $, the Hamiltonian can be written in the sphere coordinate as \cite{haldane_hierarchy,jainbook,Greiter11} \be H=\frac{{\mathbf \Lambda}^2 }{2m_e R^2},\ee where the square of the dynamical angular momentum $\mathbf \Lambda$ is \be{\mathbf \Lambda}^2=-\frac{1}{\sin \theta}\frac{\partial}{\partial\theta} \sin\theta\frac{\partial}{\partial\theta} -\frac{1}{\sin^2\theta}\left(\frac{\partial}{\partial\phi}-iQ\cos\theta\right)^2.\ee  The generator of rotations about the origin which commutes with the Hamiltonian is \be \mathbf L=\mathbf \Lambda+Q \mathbf{\hat r}.\ee
The  eigenstates of $H$, $L^2$ and $L_z$ in the $n$th Landau level are monopole harmonics  \cite{mono,jainbook}\be\begin{split}
\langle\mathbf r |  Y_{Q,l,m} \rangle=& N_{Q,l,m}2^{-m}\left(1-\cos \theta\right)^{\frac{m-Q}{2}}\left(1+\cos \theta\right)^{\frac{m+Q}{2}} \\& \times e^{i m \phi}P_{l-m}^{m-Q,m+Q}(\cos \theta),\end{split}\ee where the normalization  \be N_{Q,l,m}=\sqrt{\frac{(2l+1)(l-m)!(l+m)!}{4 \pi (l-Q)!(l+Q)!}},\ee$P_{l-m}^{m-Q,m+Q} $ is the Jacobi polynomial, $\theta$ and  $\phi$ are the polar and azimuthal angles on the sphere, respectively. The  eigenvalue of $L^2$  is $l(l+1)\hbar^2$, with the total angular momentum $l$ being the sum of monopole charge $Q$ and LL index $n$,   \be l=Q+n.\ee  The  eigenvalue of $L_z$ is  $m=-l,-l+1,\cdots, l-1,l.$

The $N$-particle wave function   at  $\nu=1$   can be written down in terms of spinor variables $u=\cos \frac{\theta}{2} e^{i\frac{ \phi}{2}}$ and $v=\sin \frac{\theta}{2} e^{-i \frac{\phi}{2}}$,\be \Psi_1=\prod\limits_{1\leq i \leq N} v_i^{N-1} \prod\limits_{1\leq j<k  \leq N}  (z_j-z_k),\ee where $ z=\frac{u}{v}=\cot \frac{\theta}{2}e^{i\phi} $.

Now let us consider the $ \nu=p/(2p+1) $ Jain state constructed from   CFs  filling the $ n_1,n_2,\cdots n_p $th  LLs. The number of CFs in the $ n_i $th LL is $ N_i $. The monopole charge for the $\nu=p$ integer quantum Hall effect of CFs is chosen as $Q^*$, which is different from the monopole charge $Q$ for the $\nu=p/(2p+1)$ fractional quantum Hall effect of electrons. The relation of $Q$ to $Q^*$ will be revealed in Eq. \ref{Q}.  The wave function of this state is \begin{widetext}\be\label{wf}
\Psi_1^2 \times \left| {\begin{array}{*{20}{c}}
  Y_{Q^*,Q^*+n_1,m_1}(\mathbf{r}_1)& \ldots & Y_{Q^*,Q^*+n_1,m_1}(\mathbf{r}_N)\\
   \vdots & \ddots & \vdots  \\
 Y_{Q^*,Q^*+n_1,m_{N_1}}(\mathbf{r}_1)& \cdots & Y_{Q^*,Q^*+n_1,m_{N_1}}(\mathbf{r}_N)\\ Y_{Q^*,Q^*+n_2,m_{N_1+1}}(\mathbf{r}_1)& \ldots & Y_{Q^*,Q^*+n_2,m_{N_1+1}}(\mathbf{r}_N) \\
    \vdots & \ddots & \vdots  \\
Y_{Q^*,Q^*+n_2,m_{N_1+N_2}}(\mathbf{r}_1)& \cdots & Y_{Q^*,Q^*+n_2,m_{N_1+N_2}}(\mathbf{r}_N)\\
   \vdots & \ddots & \vdots  \\
 Y_{Q^*,Q^*+n_p,m_{N}}(\mathbf{r}_1)& \cdots & Y_{Q^*,Q^*+n_p,m_{N}}(\mathbf{r}_N)
\end{array}} \right|.     \ee
In the same manner in which we expand the wave function on the disk in the previous appendix, here the wave function can be expanded as  \be\label{mas}\sum\limits_{\substack{i_1+i_2+\cdots +i_N=(N-1)N\\0\leq i_1,i_2,\cdots i_N\leq2(N-1)}} C_{i_1,i_2,\cdots i_N} \left| {\begin{array}{*{20}{c}}
  v_1^{2N-2}z_1^{i_1}Y_{Q^*,Q^*+n_1,m_1}(\mathbf{r}_1)& \ldots & v_N^{2N-2}z_N^{i_1}Y_{Q^*,Q^*+n_1,m_1}(\mathbf{r}_N)\\
   \vdots & \ddots & \vdots  \\
 v_1^{2N-2}z_1^{i_{N_1}}Y_{Q^*,Q^*+n_1,m_{N_1}}(\mathbf{r}_1)& \cdots & v_N^{2N-2}z_N^{i_{N_1}}Y_{Q^*,Q^*+n_1,m_{N_1}}(\mathbf{r}_N)\\ v_1^{2N-2}z_1^{i_{N_1+1}}Y_{Q^*,Q^*+n_2,m_{N_1+1}}(\mathbf{r}_1)& \ldots & v_N^{2N-2}z_N^{i_{N_1+1}}Y_{Q^*,Q^*+n_2,m_{N_1+1}}(\mathbf{r}_N) \\
    \vdots & \ddots & \vdots  \\
v_1^{2N-2}z_1^{i_{N_1+N_2}}Y_{Q^*,Q^*+n_2,m_{N_1+N_2}}(\mathbf{r}_1)& \cdots & v_N^{2N-2}z_N^{i_{N_1+N_2}}Y_{Q^*,Q^*+n_2,m_{N_1+N_2}}(\mathbf{r}_N)\\
   \vdots & \ddots & \vdots  \\
 v_1^{2N-2}z_1^{i_{N}}Y_{Q^*,Q^*+n_p,m_{N}}(\mathbf{r}_1)& \cdots & v_N^{2N-2}z_N^{i_{N}}Y_{Q^*,Q^*+n_p,m_{N}}(\mathbf{r}_N)
\end{array}} \right|, \ee
where $C_{i_1,i_2,\cdots i_N}$ is the same as that in the previous appendix.
Note that each entry in the above Slater determinant is of the form $ v^{2N-2}z^j Y_{Q^*,Q^*+n,m} $ with $0\leq j \leq 2N-2$. In order to project the wave function to LLL,  we need to simplify  each entry as a combination of monopole harmonics. Observe that   \be\label{sh}\begin{split}
v^{2N-2}z^j Y_{Q^*,Q^*+n,m}=& N_{Q^*,Q^*+n,m}2^{1-m-N}\left(1-\cos \theta\right)^{\frac{m-Q^*+2N-2-j}{2}}\left(1+\cos \theta\right)^{\frac{m+Q^*+j}{2}} \\& \times e^{i \phi (m-N+1+j)}P_{Q^*+n-m}^{m-Q^*,m+Q^*}(\cos \theta).\end{split}\ee In order to bring the above to the form of monopole harmonics, we define new monopole charge $Q$ and new $L_z$ angular momentum  $m'$,
\be\label{Q}\begin{split}
& Q=Q^*+N-1,\\& m'=m-N+1+j.
\end{split}\ee
Using this new definition, Eq. \ref{sh} can be written as \be\begin{split}
v^{2N-2}z^j Y_{Q^*,Q^*+n,m}=& N_{Q^*,Q^*+n,m}2^{1-m-N}\left(1-\cos \theta\right)^{\frac{m'-Q}{2}}\left(1+\cos \theta\right)^{\frac{m'+Q}{2}}e^{i \phi m'} \\& \times  \left(1-\cos \theta\right)^{2N-2-j}P_{Q^*+n-m}^{m-Q^*,m+Q^*}(\cos \theta).\end{split}\ee Thus we need to bring $ \left(1-\cos \theta\right)^{2N-2-j}P_{Q^*+n-m}^{m-Q^*,m+Q^*}(\cos \theta) $ to the form of $ P_{Q +n' }^{m-Q^*-2N+2+j,m+Q^*+j}(\cos \theta) $, that is, to lower the  upper left index of $ P_{Q^*+n-m}^{m-Q^*,m+Q^*}(\cos \theta) $ by $2N-2-j$ and raise its   upper right index  by $j$.
Now with a recursive formula for Jacobi polynomials \cite{tableinte}  \be (2n+\alpha+\beta) P_{n }^{\alpha,\beta-1}(x)=(n+\alpha+\beta) P_{n }^{\alpha,\beta}(x)+(n+\alpha) P_{n-1 }^{\alpha,\beta}(x),\ee
we can prove that  \be P_{Q^*+n-m}^{m-Q^*,m+Q^*}(\cos \theta)=\sum\limits_{k=0}^{j}d_{j,k}P_{Q^*+n-m-k}^{m-Q^*,m+Q^*+j}(\cos \theta),\ee
where $ d_{j,k} $(which depends on $Q^*,n$ and $m$) can be obtained   recursively,\be\begin{split}
& d_{0,0}=1,\\& d_{j,k}=\frac{Q^*+n+m-k+j}{2(Q^*+n)-2k+j}d_{j-1,k}+\frac{ n -k+1}{2(Q^*+n)-2k+2+j} d_{j-1,k-1}\quad \text{for} \quad j\geq 2, \quad 1\leq k \leq j-1,\\& d_{j,0}=\frac{Q^*+n+m+j}{2(Q^*+n)+j}d_{j-1,0},\\& d_{j,j}=\frac{ n -j+1}{2(Q^*+n)-j+2}d_{j-1,j-1}.
\end{split}\ee
With another recursive formula for Jacobi polynomials \cite{tableinte}  \be ( n+\alpha/2+\beta/2+1)(1-x) P_{n }^{\alpha+1,\beta}(x)=(n+\alpha+1) P_{n }^{\alpha,\beta}(x)-(n+1) P_{n+1 }^{\alpha,\beta}(x),\ee
we obtain \be \left(1-\cos \theta\right)^{2N-2-j}P_{Q^*+n-m}^{m-Q^*,m+Q^*}(\cos \theta)=\sum\limits_{k=0}^{j}d_{j,k}\sum\limits_{k'=0}^{2N-2-j}e_{2N-2-j,k'}P_{Q^*+n-m-k+k'}^{m-Q^*-2N+2+j,m+Q^*+j}(\cos \theta),\ee
where $ e_{2N-2-j,k'} $(which not only depends on $Q^*,n$ and $m$, but also on $k$) can also be obtained   recursively, \be\label{e}\begin{split}
& e_{0,0}=1,\\& e_{t',k'}=\frac{ (n -k+k'+1-t')e_{t'-1,k'}}{ Q^*+n-k+k'+1+j/2-t'/2}-\frac{(Q^*+n-m -k+k')e_{t'-1,k'-1}}{ Q^*+n -k+k'+j/2-t'/2} \quad \text{for}  \quad  t'\geq 2, \quad 1\leq k' \leq t'-1,\\& e_{t',0}=\frac{ n -k +1-t'}{ Q^*+n-k +1+j/2-t'/2}e_{t'-1,0},\\& e_{t',t'}=-\frac{Q^*+n-m -k+t'}{ Q^*+n -k+j/2+t'/2}e_{t'-1,t'-1}.
\end{split}\ee

Finally we have \be\label{spherical}
v^{2N-2}z^j Y_{Q^*,Q^*+n,m}=  N_{Q^*,Q^*+n,m}2^{1+j-N}\sum\limits_{k=0}^{j}\sum\limits_{k'=0}^{2N-2-j} \frac{d_{j,k}e_{2N-2-j,k'}}{N_{Q,Q+n',m'}}Y_{Q,Q+n',m'},\ee
\end{widetext}
where  $Q$ and  $m'$ are defined in Eq. \ref{Q}. New LL index $n'$ is related to old LL index $n$ by \be\label{newLL} n'=n-k-(2N-2-j-k').\ee Since $k\geq0,\, k'\leq 2N-2-j $ as seen from summation indices, $n'\leq n$ always holds, which is as expected.

Below we will show  that on the sphere we recover the same constraint on LL indices as the case on the disk when the thermodynamic limit is taken.

In the thermodynamic limit $N \rightarrow \infty$, the monopole strength $Q$ and $Q^*$ also go to   infinity. While the magnetic field on sphere $\frac{\hbar Q}{eR^2}$ is held constant, the sphere radius $R$  goes to  infinity. Consequently, the sphere is locally equivalent to the disk in this limit. It is easy to see that in the limit $Q^* \rightarrow \infty$, the $e_{t',k'}$ in Eq. \ref{e} will vanish unless  $k'=t'$. Therefore, $k'$ only take the value  $2N-2-j$ in Eq. \ref{spherical}. The new LL index $n'$ in Eq. \ref{newLL} is thus $n-k$, which is the same as that in Eq. \ref{mat2} on the disk. It then follows that in the thermodynamic limit, for the $ \nu=p/(2p+1) $ Jain state on the sphere to  have nonvanishing LLL projection, the constraints on indices of filled CF LLs would be the same as those in the case of the disk as given in Eq. \ref{constr0}.

By contrast,  when  the particle number  and monopole charge are finite, the new LL index $n'$ is given by Eq. \ref{newLL}.  The minimum of $n'$ in Eq. \ref{newLL} must be nonpositive in order for each entry in the  Slater determinant expansion of the wave function in Eq. \ref{ma} to have nonvanishing LLL projection. In that case, we  must have the following constraints on LL indices $n_1,n_2,\dotsc, n_p$, \be\  n_1,n_2,\dotsc, n_p\leq 2N-2.\ee


Note that the constraint on LL indices in the case of finite particle number  allows more choices than those in the thermodynamic limit.

In conclusion, the   necessary conditions for general $ \nu=p/(2p+1) $ CF Jain states on  the sphere    constructed from the $  n_1,n_2,\dotsc, n_p $th CF LLs  filled with $  N_1,N_2,\dotsc, N_p$ CFs($N=\sum_{i=1}^p N_i$) to have nonvanishing LLL projection in the thermodynamic limit are (1)  $ \sum_{i=1}^p N_i n_i\leq (N-1)N $  and (2) the maximum of LL indices  is no  greater than $2(N-1)$. When the particle number $N$ is finite, we only have the second constraint on CF  LL indices.

\bibliography{cft}
\end{document}